\documentclass[conference]{IEEEtran}
\IEEEoverridecommandlockouts
\usepackage{cite}
\usepackage{amsmath,amssymb,amsfonts}
\usepackage{algorithmic}
\usepackage{graphicx}
\usepackage{textcomp}
\usepackage{xcolor}

\usepackage[justification=centering]{caption}
\usepackage[labelformat=simple]{subcaption}

\usepackage{booktabs}
\usepackage{makecell}

\renewcommand{\vec}[1]{\boldsymbol{#1}}

\def\BibTeX{{\rm B\kern-.05em{\sc i\kern-.025em b}\kern-.08em
    T\kern-.1667em\lower.7ex\hbox{E}\kern-.125emX}}
\begin{document}

\title{
Refined WaveNet Vocoder for Variational Autoencoder Based Voice Conversion
}

\author{\IEEEauthorblockN{
	Wen-Chin Huang\IEEEauthorrefmark{1},
	Yi-Chiao Wu\IEEEauthorrefmark{2},
	Hsin-Te Hwang\IEEEauthorrefmark{1},
	Patrick Lumban Tobing\IEEEauthorrefmark{2},
	Tomoki Hayashi\IEEEauthorrefmark{2}, \\
	Kazuhiro Kobayashi\IEEEauthorrefmark{2},
	Tomoki Toda\IEEEauthorrefmark{2},
	Yu Tsao\IEEEauthorrefmark{1},
	Hsin-Min Wang\IEEEauthorrefmark{1}
	}
	\IEEEauthorblockA{\IEEEauthorrefmark{1}
	Academia Sinica, Taiwan}
	\IEEEauthorblockA{\IEEEauthorrefmark{2}Nagoya University, Japan}
	}

\maketitle

\begin{abstract}
This paper presents a refinement framework of WaveNet vocoders for variational autoencoder (VAE) based voice conversion (VC), which reduces the quality distortion caused by the mismatch between the training data and testing data. Conventional WaveNet vocoders are trained with natural acoustic features but conditioned on the converted features in the conversion stage for VC, and such a mismatch often causes significant quality and similarity degradation. In this work, we take advantage of the particular structure of VAEs to refine WaveNet vocoders with the self-reconstructed features generated by VAE, which are of similar characteristics with the converted features while having the same temporal structure with the target natural features. We analyze these features and show that the self-reconstructed features are similar to the converted features. Objective and subjective experimental results demonstrate the effectiveness of our proposed framework.
\end{abstract}


\section{Introduction}
Voice conversion (VC) aims to convert the speech from a source to that of a target without changing the linguistic content. While numerous approaches have been proposed \cite{661472,4317579, 5445041,6891242, 6424242,6843941,Wu+2016}, in this work we study variational autoencoder \cite{VAE} based VC (VAE-VC) \cite{VAEVC}, where the conversion framework consists of an encoder-decoder pair. First, the source features are encoded into a latent content code by the encoder, and the decoder generates the converted features by conditioning on the latent code and the target speaker code. The whole network is trained in an unsupervised manner, minimizing the reconstruction error and a Kullback-Leibler (KL)-divergence loss, which regularizes the distribution of the latent variable. Because of the unsupervised learning nature of VAE, VAE-VC does not need parallel training data.

\begin{figure}[t]
	\centering
	\includegraphics[width=0.45\textwidth]{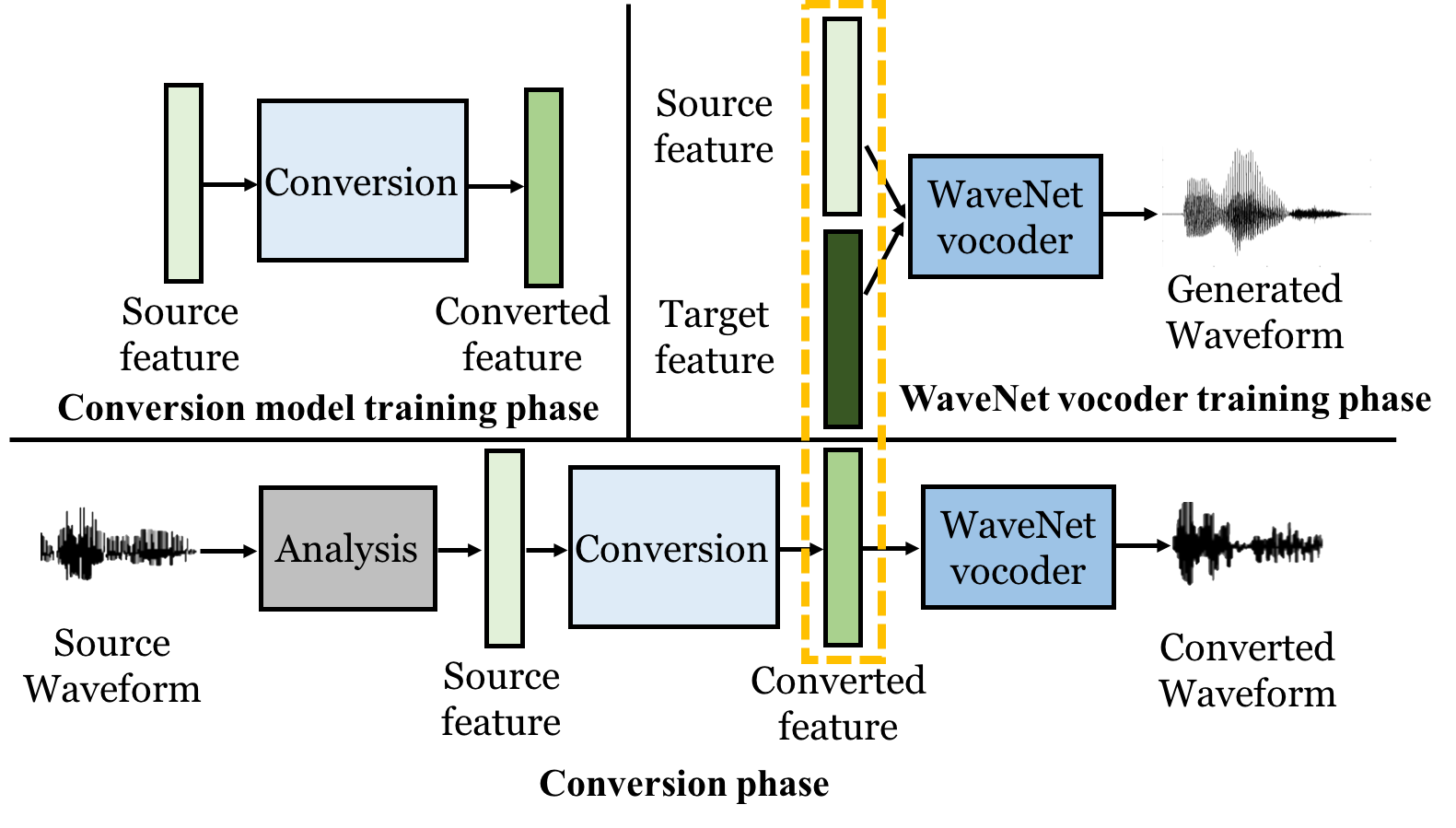} 
	\caption{A general mismatch between the training and conversion features of the WaveNet vocoder, which is highlighted in the orange box.	\label{fig:mismatch}}
\end{figure}

The waveform generation process plays an important role in a VC system. Conventional VC frameworks employ parametric vocoders as their synthesis module, which impose many overly simplified assumptions that discard the phase information and result in unnatural excitation signals, and thus cause a significant degradation in the quality of the converted speech \cite{LPC, STRAIGHT, WORLD}. In recent years, the WaveNet vocoder \cite{SDWN, SIWN}, built upon one of the most promising neural speech generation models, WaveNet \cite{wavenet}, has been proposed. It is capable of reconstructing the phase and excitation information, and thus generates more natural sounding speech.

In \cite{SIWN}, it was confirmed that compared with the speaker independent (SI) WaveNet vocoder, which is trained with a multi-speaker speech corpus, the speaker dependent (SD) variant \cite{SDWN} is superior in terms of speech quality. However, it is impractical to collect a sufficient amount of data (usually more than 1 hour) required to build a target-dependent WaveNet vocoder in a VC framework. To alleviate this problem, many researchers have applied a fine-tuning technique \cite{NU-VCC2018-P,NU-VCC2018-NP,SAWN-Sisman,SAWN-ROCLING,SAWN-iFLYTEK,cyclernn} and its effectiveness has been confirmed in \cite{SAWN-Sisman}.

\begin{figure*}[t]

\begin{minipage}[b]{0.4\textwidth}
  \centering
  \includegraphics[width=\textwidth]{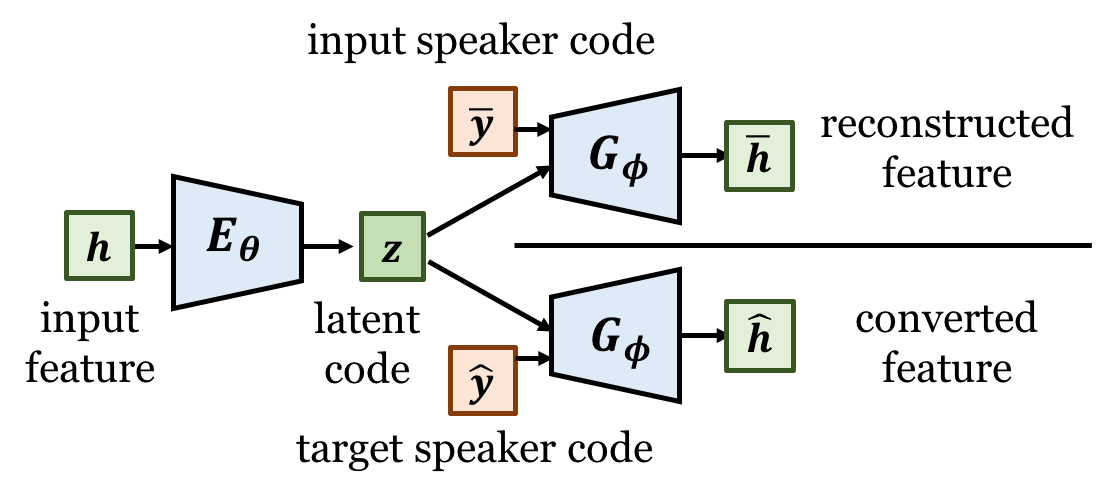}
  \captionof{figure}{Illustration of the two types of spectral frames generated by feeding different speaker codes in the forward process of VAE-VC.}
  \label{fig:vae-flow}
\end{minipage}
~
\begin{minipage}[b]{0.6\textwidth}
  \centering
  \includegraphics[width=0.85\textwidth]{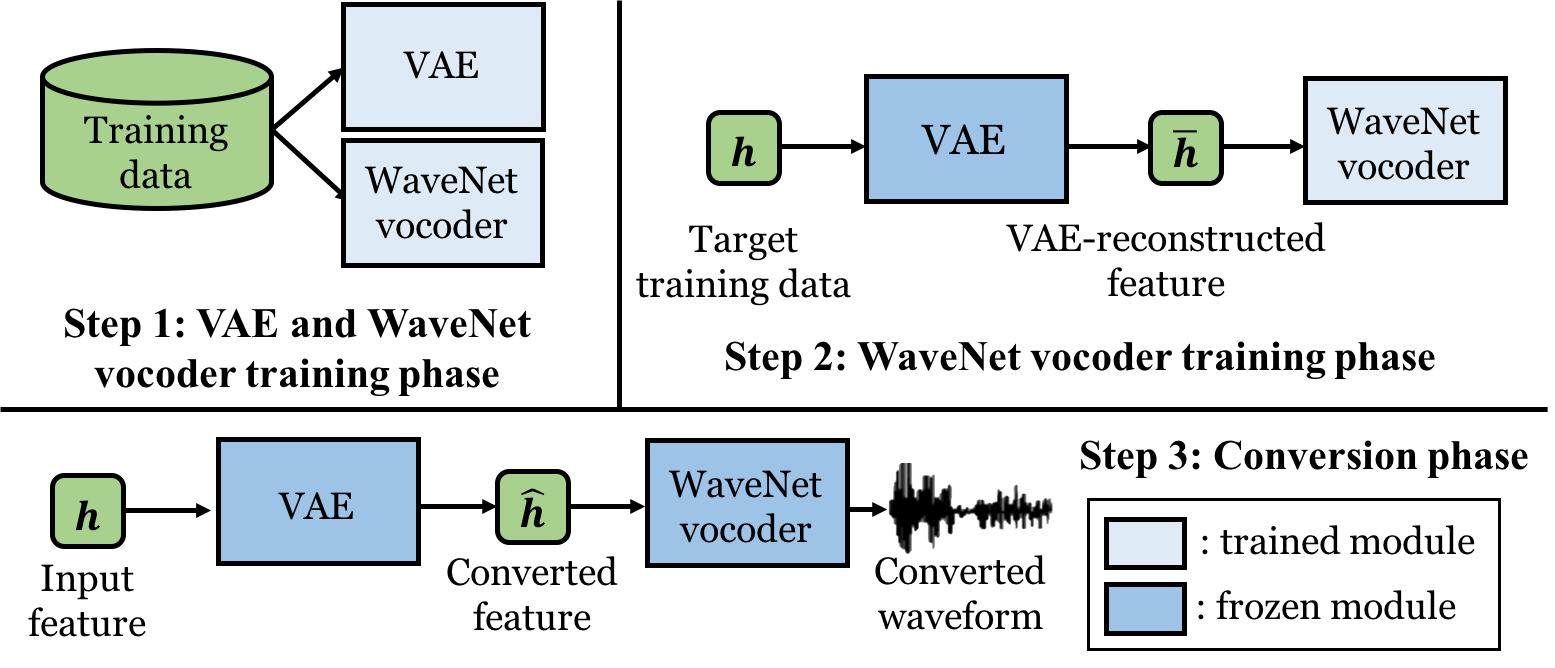}
  \captionof{figure}{The proposed WaveNet vocoder fine-tuning framework using \textit{VAE-reconstructed} features. The analysis module is omitted.}
  \label{fig:proposed}
\end{minipage}
\end{figure*}

However, the above mentioned fine-tuning technique suffers from the mismatch between the training and testing input features of the WaveNet vocoder, as shown in Fig.~\ref{fig:mismatch}. The features used to train the WaveNet vocoder are extracted from either the source's or the target's natural speech, which we will refer to as the \textit{natural} features, and are different from the \textit{converted} features generated by the conversion model. One way to reduce this mismatch is to use the converted features to fine-tune the WaveNet vocoder, similar to \cite{taco2}. However, this cannot be used directly in VC, since the time alignment between the converted features and the target speech waveform will be required, and an imperfect alignment will introduce an additional mismatch.

To generate alignment-free data, \cite{VCWN} first used intra-speaker conversion, but the performance was limited since there was a gap between such intra-converted features and the actual converted features. Furthermore, \cite{cyclernn} used a cyclic  conversion model to ensure the self-predicted features were indeed similar to the converted features. Although these methods seem to be promising in parallel VC, the use of such techniques in non-parallel VC has not been properly investigated. For instance, it remains unknown how to effectively apply such techniques to VAE-VC.

In light of this, in this work, we propose a novel WaveNet vocoder fine-tuning framework for VAE-VC. Specifically, the self-reconstructed features of the trained VAE network are used to fine-tune the WaveNet vocoder. Our contributions are:
\begin{itemize}
  \item The self-reconstructed features of VAE are suitable for fine-tuning. Since the forward processes of the VAE-VC are almost identical except that they are conditioned on different speaker codes, we hypothesize that the characteristics of the self-reconstructed features are similar to that of the converted features. We analyzed the properties of these features and verified this hypothesis with objective measures. To our knowledge, this is the first attempt to examine the "suitableness" of the features used for fine-tuning.
  \item The self-reconstructed features of VAE are innately time-alignment free, since the temporal structure of self-reconstructed and natural target features is identical. Therefore, we don't need to add additional networks or losses to solve the alignment issue, as in \cite{VCWN, cyclernn}. 
\end{itemize}

\section{Related techniques}
\label{sec:related}

\subsection{VAE-VC}
\label{ssec:VAEVC}

In VAE-VC \cite{VAEVC}, the conversion function is formulated as an encoder-decoder network. Specifically, given an observed (source or target) spectral frame $\vec{h}$, a speaker-independent encoder $E_\theta$ with parameter set $\theta$ encodes $\vec{h}$ into a latent code: $\vec{z}=E_\theta(\vec{h})$. The (target) speaker code $\vec{\hat{y}}$ is then concatenated with the latent code, and passed to a conditional decoder $G_\phi$ with parameter set $\phi$ to generate the \textit{converted} features. Thus, the conversion function $f$ of VAE-VC can be expressed as: $\hat{\vec{h}}=f(\vec{h})=G_\phi(\vec{z},\vec{\hat{y}})$. Fig.~\ref{fig:vae-flow} shows the two types of spectral frames used in training and conversion. In the training phase, the forward pass is done by encoding $\vec{h}$ of an arbitrary speaker into a latent code $\vec{z}$, and feeding it into the decoder along with the speaker code $\bar{\vec{y}}$ to generate the reconstructed frame $\bar{\vec{h}}$, which we will refer to as the \textit{VAE-reconstructed} feature.

During training, there is no need for parallel training data. The model parameters can be obtained by maximizing the variational lower bound, $\mathcal{L}$:
\begin{equation}
\mathcal{L}(\theta,\phi;\vec{h}, \vec{y}) = \mathcal{L}_{recon}(\vec{h},\vec{y})+\mathcal{L}_{lat}(\vec{h}),
\end{equation}
\begin{equation}
\mathcal{L}_{recon}(\vec{h},\vec{y})=\mathbb{E}_{q_\theta(\vec{z}|\vec{h})}\bigl[\log p_\phi(\vec{h}|\vec{z},\vec{y})\bigr],
\label{recon_loss}
\end{equation}
\begin{equation}
\mathcal{L}_{lat}(\vec{h})=-D_{KL}(q_\theta(\vec{z}|\vec{h}) \Vert p(\vec{z})),
\label{lat_loss}
\end{equation}
where $q_\theta(\vec{z}|\vec{h})$ is the approximate posterior, $p_\phi(\vec{h}|\vec{z},\vec{y})$ is the data likelihood, and $p(\vec{z})$ is the prior distribution of the latent space. $\mathcal{L}_{recon}$ is simply a reconstruction term as in any vanilla autoencoder, whereas $\mathcal{L}_{lat}$ regularizes the encoder to align the approximate posterior with the prior distribution.

\begin{figure*}[t]
\begin{minipage}[b]{0.28\textwidth}
  \centering
  \includegraphics[width=0.8\textwidth]{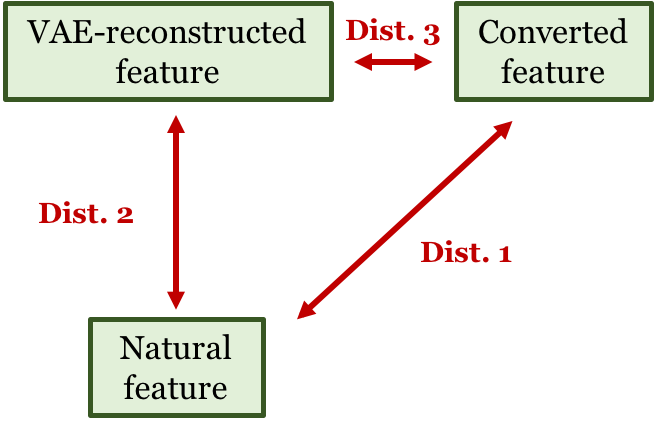}
  \captionof{figure}{An illustration of the distances between the three types of features.}
  \label{fig:analysis-distance}
\end{minipage}
~
\begin{minipage}[b]{0.36\textwidth}
  \centering
  \includegraphics[width=0.9\textwidth]{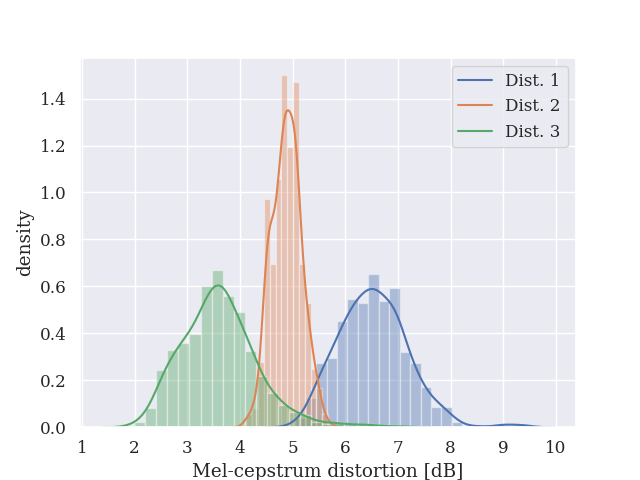}
  \captionof{figure}{Density plot of the mel-cepstral distortion of the three distances.}
  \label{fig:analysis-MCD}
\end{minipage}
~
\begin{minipage}[b]{0.34\textwidth}
  \centering
  \includegraphics[width=0.9\textwidth]{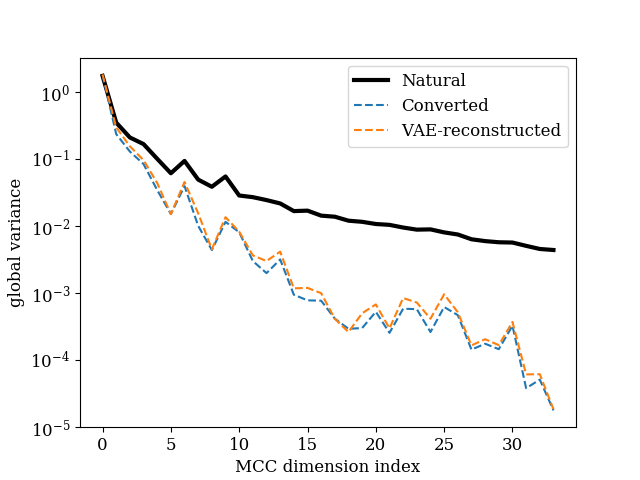}
  \captionof{figure}{Global variance values of the three types of features from speaker TF1.}
  \label{fig:analysis-gv}
  
\end{minipage}
\end{figure*}

\subsection{WaveNet vocoder}
\label{ssec:WN}

WaveNet \cite{wavenet} is a deep autoregressive network, which generates high-fidelity speech waveforms sample-by-sample, using the following conditional probability equation:
\begin{equation}
P(\vec{X}|\vec{h})=\prod_{n=1}^{N}P(x_n|x_{n-r},\dots,x_{n-1}, \vec{h}),
\label{eq:wn}
\end{equation}
where $x_n$ is the current sample point, $r$ is the size of the receptive field, and $\vec{h}$ is the auxiliary feature vector. 
The WaveNet vocoder \cite{SDWN,SIWN} reconstructs time-domain waveforms by conditioning on acoustic features, including spectral features, fundamental frequencies ($f_0$) and aperiodicity signals (APs). A WaveNet vocoder is composed of several stacked residual blocks skip-connected to the final output, with each residual block containing a $2\times1$ dilated causal convolution layer, a gated activation function, and two $1\times1$ convolution layers. In short, given a data sequence pair $(\vec{h}, \vec{x})$, the WaveNet vocoder is trained in a supervised fashion, learning to map the acoustic features $\vec{h}$ to the time-domain signals $\vec{x}$.

In recent years, fine-tuning a WaveNet vocoder to the target speaker has become a popular technique to improve speech quality when applying the WaveNet vocoder to VC \cite{NU-VCC2018-P,NU-VCC2018-NP,SAWN-Sisman,SAWN-ROCLING,SAWN-iFLYTEK,cyclernn}. Specifically, an initial SI WaveNet vocoder is first trained with a multi-speaker dataset. Then, the target speaker's data is used to fine-tune all or partial model parameters. Therefore, suitable fine-tuning data that matches the converted features is the key to better speech quality.

\subsection{VAE-VC with WaveNet vocoder}
\label{ssec:WN-VAE}

A combination of VAE-VC and WaveNet vocoder has been proposed in \cite{SAWN-ROCLING}. First, a VAE conversion model is trained. A WaveNet vocoder is then constructed by training with natural features of a multi-speaker dataset first, followed by fine-tuning using the natural features of the target speaker. In the conversion phase, the VAE model first performs spectral feature conversion, and by conditioning on the converted spectral features, source APs and transformed $f_0$, the WaveNet vocoder generates the converted speech waveform. Here we once again point out that the WaveNet vocoder conditions on features of different characteristics in the training and conversion phases, i.e. there is a mismatch between the input features in the training and conversion phases. In the following section, we will introduce our proposed method for reducing this mismatch.

\section{Proposed method for WaveNet vocoder fine-tuning in VAE-VC}
\label{sec:proposed}

Using the converted features to fine-tune the WaveNet vocoder is an effective approach to reduce the mismatch between training and conversion in VC. However, this is only possible when a parallel corpus is available, which conflicts with the capability of VAE-VC under non-parallel conditions. Even with parallel data, since conventional VC models perform conversion frame-by-frame, the converted feature and the source speech have the same temporal structure, which is different from that of the target. As a result, we need an elegant fine-tuning method for a non-parallel VC method like VAE-VC that solves both the time-alignment and mismatch issues.

We propose to fine-tune WaveNet vocoders with VAE-reconstructed features by taking advantage of the particular network structure of VAEs, as depicted in Fig.~\ref{fig:proposed}. Specifically, a VAE model is first trained with the whole training corpus. Then, the VAE-reconstructed features $\bar{\vec{h}}$ could be obtained from the target speaker's training data through the reconstruction process described in Section~\ref{ssec:VAEVC}. We can thereafter use data pairs $(\bar{\vec{h}}, \vec{x})$ to fine-tune the SI WaveNet vocoder instead of using the original $(\vec{h}, \vec{x})$ as described in Section~\ref{ssec:WN}, where $\vec{h}$ and $\vec{x}$ denote the natural target feature and the target waveform, respectively.

There are two main advantages of using the VAE-reconstructed features to fine-tune the SI WaveNet vocoder. First, since $\vec{h}$ and $\bar{\vec{h}}$ have the same temporal structure, time-alignment is no longer required. More importantly, the mismatch between the VAE-reconstructed features and the converted features $\hat{\vec{h}}$ is small. Recall that in VAE-VC, $\bar{\vec{h}}$ and $\hat{\vec{h}}$ can be obtained by feeding the input speaker code $\bar{\vec{y}}$ and the target speaker code $\hat{\vec{y}}$, respectively. Since $\bar{\vec{h}}$ and $\hat{\vec{h}}$ only differ in speaker codes, we hypothesize that they share similar properties. Experimental results confirm the above hypothesis, as we will show in Section~\ref{ssec:analysis-mismatch}.

\section{Experimental evaluations}
\label{sec:experiments}

\subsection{Experimental settings}
\label{ssec:settings}

All experiments were conducted using the Voice Conversion Challenge 2018 (VCC2018) corpus \cite{vcc2018}, which included recordings of 12 professional US English speakers with a sampling rate of 22050 Hz and a sample resolution of 16 bits. The training set consisted of 81 utterances and the testing set consisted of 35 utterances of each speaker. The total number of training utterances was 972 and the training data length was roughly 54 minutes. The WORLD vocoder \cite{WORLD} was adopted to extract acoustic features including 513-dimensional spectral envelopes (SPs), 513-dimensional APs and $f_0$. The SPs were normalized to unit-sum, and the normalizing factor was taken out and thus not modified. 35-dimensional mel-cepstral coefficients (MCCs) were extracted from the SPs.

For the spectral conversion model, we used the CDVAE variant \cite{CDVAE} as it was capable of modeling MCCs while the original VAE \cite{VAEVC} failed to do so \cite{CDVAE}. The CDVAE model was trained with the whole training data of the 12 speakers from VCC2018, with the detailed network architecture and training hyper-parameters consistent with \cite{CDVAE}.  

For the WaveNet vocoder, we followed the official implementation\footnote{https://github.com/kan-bayashi/PytorchWaveNetVocoder} with the default settings, including the noise shaping technique \cite{noiseshaping}. All training data of VCC2018 were used to train the multi-speaker WaveNet vocoder. Although some works used discriminate speaker embeddings for training, we did not adopt such technique since simply mixing all speakers gave better performance as reported in \cite{SIWN}. The default 200000 iterations of model training led to approximate 2.0 training loss (in terms of negative log-likelihood). Fine-tuning of the WaveNet vocoder was performed by updating the whole network until the training loss reached around 1.0. At conversion, the input features included the converted MCCs from the VAE conversion model followed by energy compensation, the linear mean-variance transformed log-$f_0$ as well as the unmodified source APs.

\subsection{Analysis of Mismatch}
\label{ssec:analysis-mismatch}

In order to prove that the VAE-reconstructed features are more similar to the converted features than the natural target features, we analyzed their spectral properties based on the mean mel-cepstral distortion (MCD) and global variance (GV) measurements on the training set. First, we trained a VAE and obtained the three types of features using the training data of the VCC2018 corpus. Then, we calculated the distances (Dist. 1-3) by calculating the mean MCD of each sentence and then comparing the distributions, as illustrated in Fig.~\ref{fig:analysis-distance}.

Fig.~\ref{fig:analysis-MCD} shows the three distances calculated from conversion pairs of a subset of speakers (SF1, SM1 to TF1, TF2, TM1, TM2). First, we observe that Dist. 2 is rather large, showing that the natural features are somehow distorted through the imperfect reconstruction process of VAE. We further observe that Dist. 3 is smaller than Dist. 1, demonstrating that conventional approaches suffer from the mismatch given by Dist. 1, and our proposed method can alleviate the issue because Dist. 3 is small. Note that Dist. 1 and Dist. 3 are affected by time alignment while Dist. 2 is not, so the former have larger variation than the latter.

Fig.~\ref{fig:analysis-gv} illustrates the GVs of the three types of features. The GVs of the natural features are higher than that of the other two types of features, since the natural features contain the most spectral details and structures while the VAE-reconstructed and converted features suffer from the over-smoothing effect. These results imply that the VAE-reconstructed features are more suitable to fine-tune the WaveNet vocoder as they well simulate the properties of the converted features.

\begin{table}[t]
	\centering
	\captionsetup{justification=centering}
	\caption{The compared methods. GV stands for the global variance postfilter \cite{fastGV}.}
	\centering
	\begin{tabular}{ c c c c c c c}
		\toprule	
		Name & Vocoder & GV & \makecell{Training\\feature} & \makecell{Adapting feature} \\
		\midrule
		Baseline 1 & WORLD & No & None & None & \\
		Baseline 2 & WORLD & Yes & None & None & \\
		Baseline 3 & WaveNet & No & \textit{Natural} & \textit{Natural} & \\
		Baseline 4 & WaveNet & Yes & \textit{Natural} & \textit{Natural} & \\
		\midrule
		Proposed 1 & WaveNet & No & \textit{Natural} & \textit{VAE-reconstructed} & \\
		Proposed 2 & WaveNet & Yes & \textit{Natural} & \textit{VAE-reconstructed}+GV & \\
		\midrule
		Upper bound & WaveNet & -- & \textit{Natural} & \textit{Natural} \\
		\bottomrule
	\end{tabular}
	\label{tab:methods}
\end{table}

\begin{figure}[t]
	\centering
	
	\begin{subfigure}[b]{0.48\textwidth}
		\centering
		\includegraphics[width=0.72\textwidth]{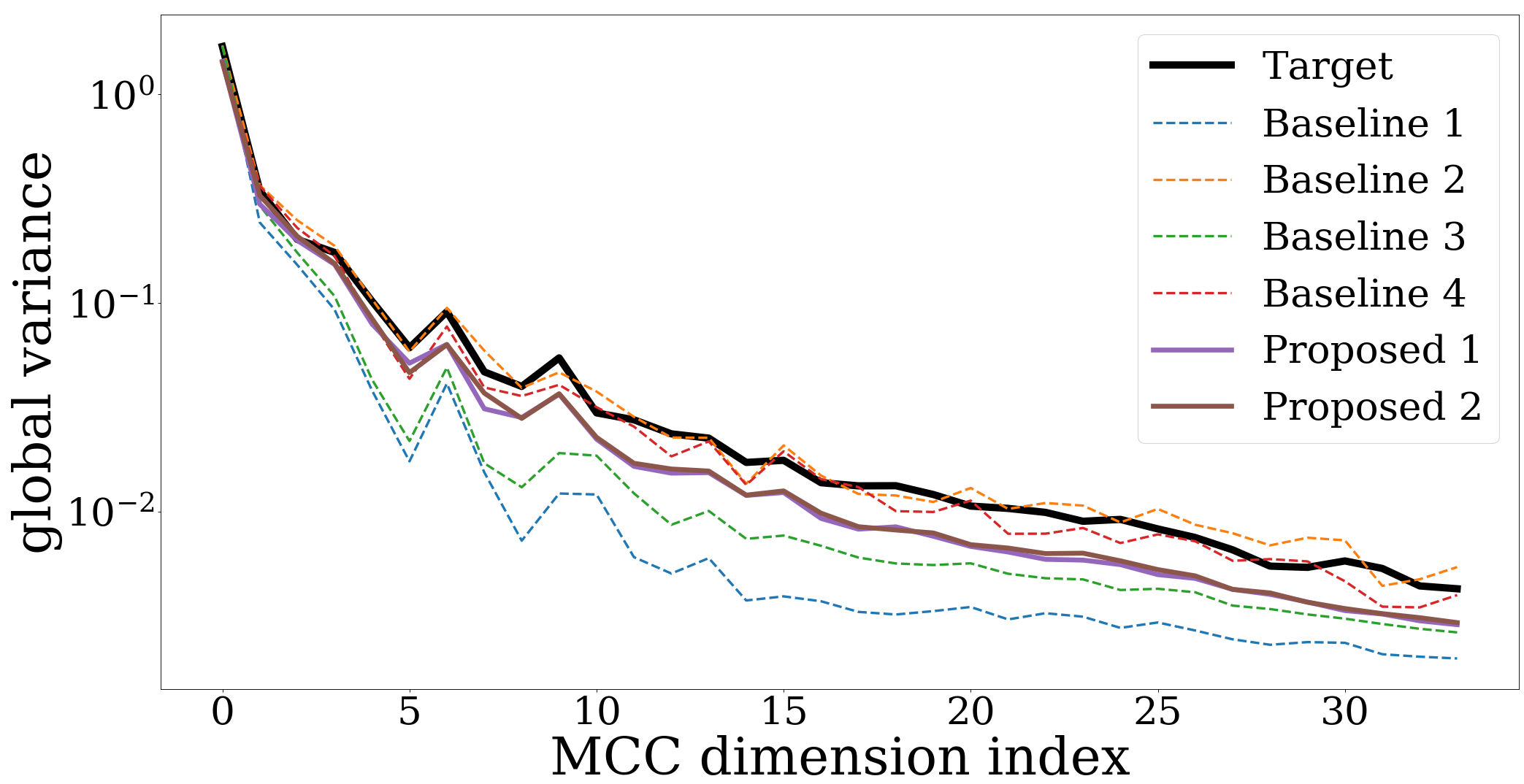} 
		\caption{Global variance values of mel-cepstrum coefficients extracted from speech converted to speaker TF1 and the natural TF1 speech. \label{fig:evaluation-gv}}
	\end{subfigure}\\
	
	\begin{subfigure}[b]{0.5\textwidth}
		\centering
		\includegraphics[width=0.75\textwidth]{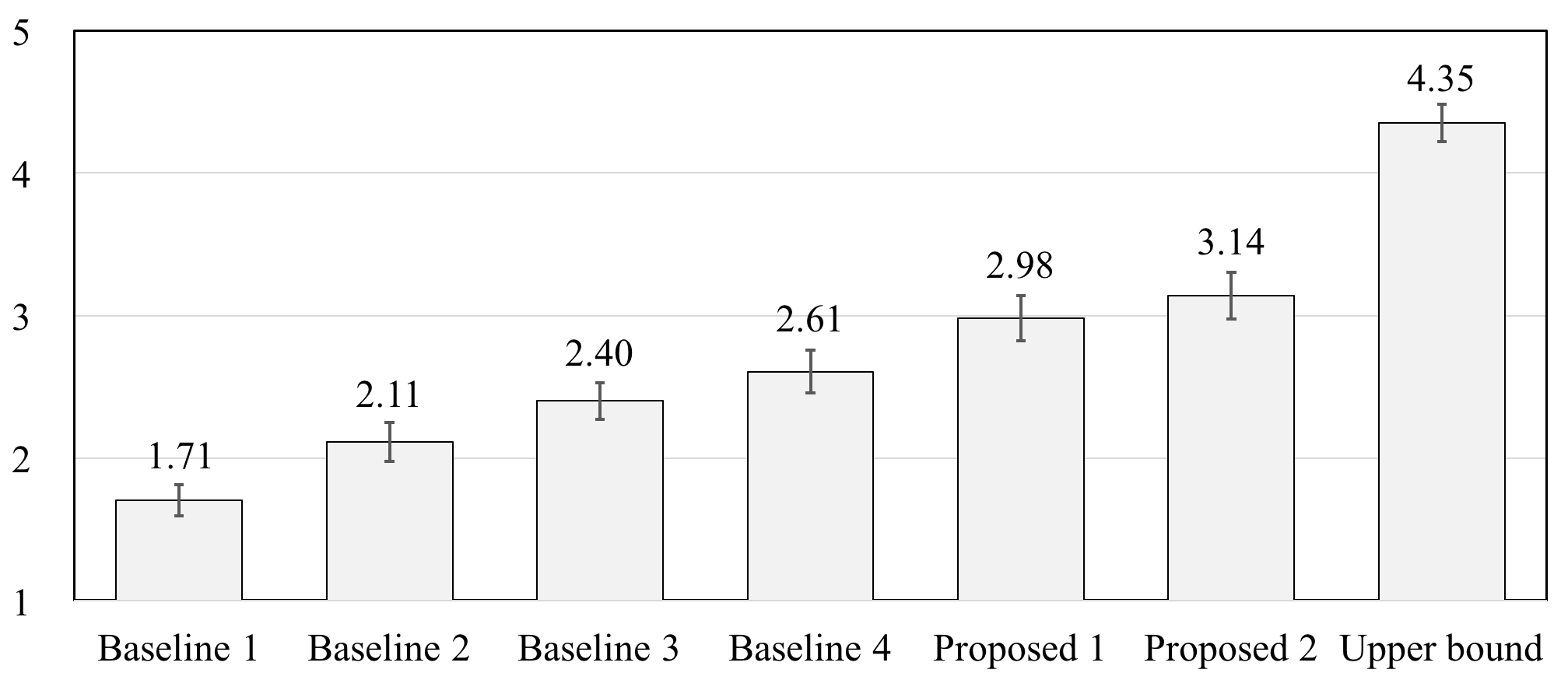} 
		\caption{Mean opinion score on naturalness over all speaker pairs. Error bars indicate the 95\% confidence intervals. \label{fig:mos}}
	\end{subfigure}\\
	
	
	\begin{subfigure}[b]{0.5\textwidth}
		\centering
		\includegraphics[width=0.75\textwidth]{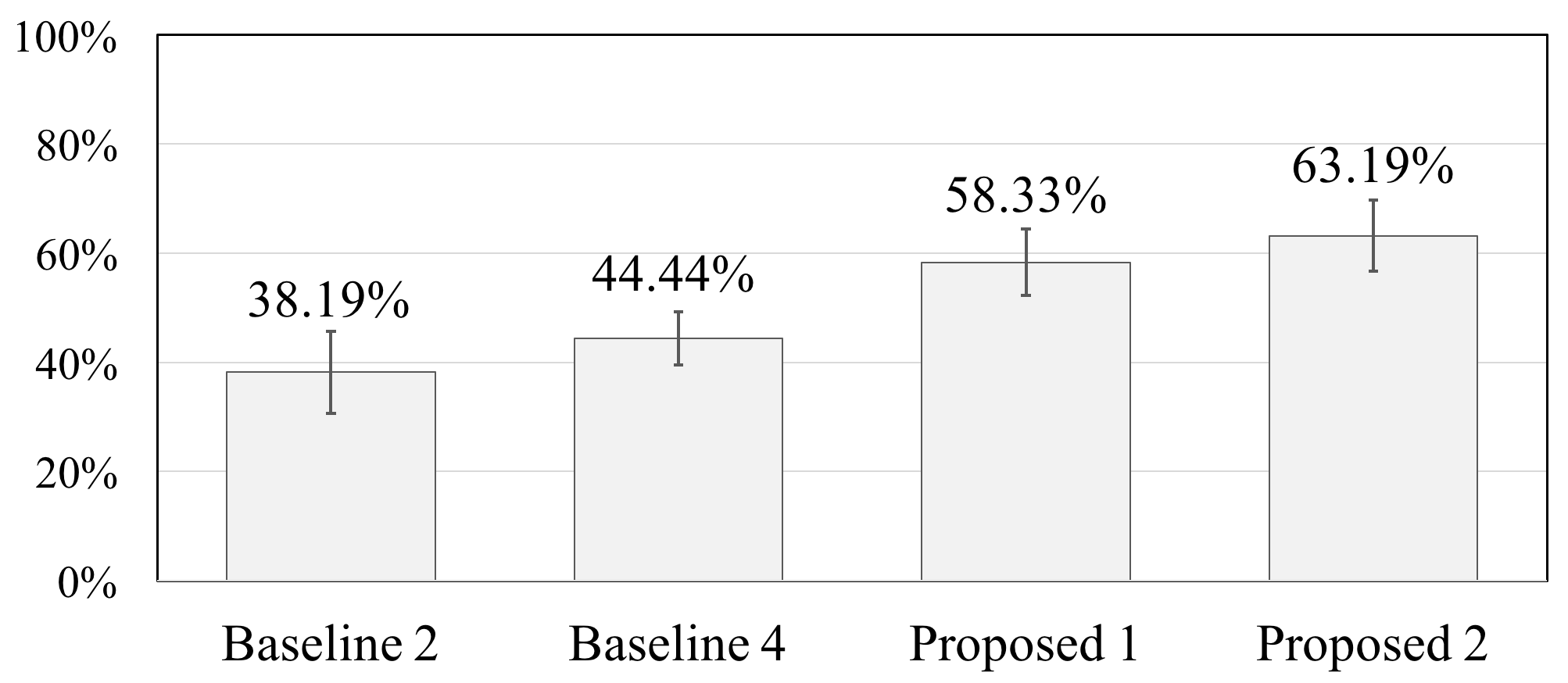} 
		\caption{Result of speaker similarity scores, which were aggregated from ``same sure" and ``same not-sure" decisions. Error bars indicate the 95\% confidence intervals.\label{fig:similarity}}
	\end{subfigure}
	
	\centering
	\caption{Results of objective and and subjective evaluations.	\label{fig:subjective}}
\end{figure}

\subsection{Evaluation of the proposed methods}
\label{ssec:eval}

We conducted objective and subjective evaluations of seven methods, as shown in Table~\ref{tab:methods}. The baseline methods (Baseline 1-4) are the VAE-VC systems with the WORLD and the fine-tuned WaveNet vocoders using the natural features \cite{CDVAE, SAWN-ROCLING}. The proposed methods (Proposed 1-2) are the VAE-VC systems with the WaveNet vocoder fine-tuned with the VAE-reconstructed and GV post-filtered VAE-reconstructed features, respectively. The upper bound system is the WaveNet vocoder fine-tuned with the target natural features, i.e. without the conversion process. Note that the GV column indicates whether the GV post-filter is applied to the converted spectral features before sending them into the WaveNet vocoder.

Fig.~\ref{fig:evaluation-gv} shows the GV over the MCCs extracted from the converted voices of the compared methods. In the VC literature, the GV values reflect perceptual quality and intelligibility, and we generally seek for a GV curve close to that of the target. In light of this, the following are some observations from the graph:
\begin{itemize}
  \item Baselines 2 and 4 had GV very close to that of the natural target, compared to Baselines 1 and 3, which is consistent with the results of combining the GV post-filtering method with the WaveNet vocoder as in \cite{VCWN}.
  \item The GV of Baseline 3 was far from that of the natural target compared to that of Baseline 4, possibly because the post-filtered converted features are closer to the natural features used to train/fine-tune the WaveNet vocoder.
  \item Proposed 1 had higher GV values than Baseline 3. This indicates that the WaveNet vocoder acts like a post-filter that compensates the over-smoothed converted features in our proposed framework.
  \item Nonetheless, the GV values of Proposed 1 and Proposed 2 were nearly identical and still lower than that of the natural target. We speculate that since we penalize WaveNet by training it using distorted features, degradation in speech quality might still be introduced.
\end{itemize}


Subjective tests were conducted to evaluate the naturalness and similarity. A 5-scaled mean opinion score (MOS) test was performed to assess the naturalness. For the speaker similarity, each listener was given a pair of audio stimuli, consisting of a natural speech of a target speaker and a converted speech, and asked to judge whether they were produced by the same speaker, with a confidence decision, i.e., sure and not sure. A subset of speakers (SF1, SM1, SF3, SM3, TF1, TM1, TF2, TM2) were chosen. Ten subjects were recruited. 

Fig.~\ref{fig:mos} shows that our proposed methods outperformed all baseline systems, even though the proposed methods had smaller GV values compared with Baselines 2 and 4. It could be inferred that the proposed methods not only alleviated the over-smoothing problem of the converted features but also improved other aspects of the generated speech, which might correspond to the naturalness of speech. Furthermore, Fig.~\ref{fig:similarity} indicates that our proposed methods also improved the speaker similarity of the generated speech.

\section{Conclusion}
\label{sec:conclusion}

In this work, we have proposed a refinement framework of VAE-VC with the WaveNet vocoder. We utilize the self-reconstruction procedure in the VAE-VC framework to fine-tune the WaveNet vocoder to alleviate the performance degradation issue caused by the mismatch between the training phase and conversion phase of the WaveNet vocoder. Evaluation results show the effectiveness of the proposed method in terms of naturalness and speaker similarity. In the future, we plan to investigate the use of VAE-reconstructed features in the multi-speaker training stage of the WaveNet vocoder to further improve the robustness. Speech samples are available at https://unilight.github.io/VAE-WNV-VC-Demo/

\section{Acknowledgement}
This work was partly supported by JST, PRESTO Grant Number JPMJPR1657 and JSPS KAKENHI Grant Number JP17H06101, as well as the MOST-Taiwan Grants 105-2221-E001-012-MY3 and 107-2221-E-001-008-MY3.

\bibliographystyle{IEEEbib}
\bibliography{strings} 

\end{document}